\documentclass[twocolumn]{aastex7}

\newcommand{\Msun}{\,{\rm M_\odot}}

\newcommand{\Mblack}{M_\bullet}

\shorttitle{Little Red Dots Are Nurseries of Massive Black Holes}
\shortauthors{Pacucci, Hernquist \& Fujii}

\begin{document}

\title{Little Red Dots Are Nurseries of Massive Black Holes}

\correspondingauthor{Fabio Pacucci} 

\author[orcid=0000-0001-9879-7780]{Fabio Pacucci}
\affiliation{Center for Astrophysics $\vert$ Harvard \& Smithsonian, 60 Garden St, Cambridge, MA 02138, USA}
\affiliation{Black Hole Initiative, Harvard University, 20 Garden St, Cambridge, MA 02138, USA}
\email[show]{fabio.pacucci@cfa.harvard.edu}

\author[0000-0001-6950-1629]{Lars Hernquist}
\affiliation{Center for Astrophysics $\vert$ Harvard \& Smithsonian, 60 Garden St, Cambridge, MA 02138, USA}
\email{lhernquist@cfa.harvard.edu}
\affiliation{Astronomy Department, Harvard University, 60 Garden St, Cambridge, MA 02138, USA}

\author[0000-0002-6465-2978]{Michiko Fujii}
\affiliation{Department of Astronomy, The University of Tokyo, 7-3-1 Hongo, Bunkyo-ku, Tokyo 113-0033, Japan}
\email{fujii@astron.s.u-tokyo.ac.jp}

\begin{abstract}
The James Webb Space Telescope (JWST) has revealed a previously unknown population of compact, red galaxies at $z \sim 5$, known as ``Little Red Dots'' (LRDs). With effective radii of $\sim 100$ pc and stellar masses of $10^9-10^{11} \Msun$, a purely stellar interpretation implies extreme central densities, $\rho_\star\sim10^4-10^5 \Msun \, \mathrm{pc}^{-3}$ and in some cases up to $\sim 10^9 \Msun \, \mathrm{pc}^{-3}$, far exceeding those of globular clusters. At such densities, the dynamical friction time for $10 \Msun$ stars in the central $0.1$ pc is $< 0.1$ Myr, driving rapid mass segregation. We investigate the dynamical consequences of such an environment using: (i) a Fokker-Planck analysis of long-term core evolution, (ii) an analytical model for the collisional growth of a very massive star (VMS), and (iii) direct $N$-body simulations. All approaches show that runaway collisions produce a VMS with mass $9\times10^3  < M_{\rm VMS} \, [\Msun] < 5\times10^4$ within $<1$ Myr. Once the supply of massive stars is depleted, the VMS contracts on a $\sim 8000$ yr Kelvin-Helmholtz timescale and undergoes a general relativistic collapse, leaving a massive black hole of mass $\Mblack \sim 10^4 \Msun$. We conclude that LRDs are natural nurseries for the formation of heavy black hole seeds via stellar-dynamical processes. This pathway produces seed number densities that far exceed those expected from direct collapse models, and, owing to the dense residual stellar core, can sustain high rates of tidal disruption events.
\end{abstract}



\section{Introduction}
\label{sec:introduction}
Early observations by the James Webb Space Telescope (JWST) revealed that the newly uncovered faint infrared sky was covered by compact and red galaxies: the Little Red Dots (LRDs, \citealt{Kocevski_2023, Harikane_2023, Matthee_2023}).
LRDs soon began confronting our understanding of the distant Universe: early studies showed that they may be very massive galaxies \citep{Labbe_2023}, possibly even ``breaking'' the $\Lambda$CDM standard cosmological model \citep{BK_2023}.

Collections of LRDs, both photometric and spectroscopic, showed that they are numerous, with number densities at $z \sim 5$ intermediate between those of UV-selected AGN and standard galaxies \citep{Kocevski_2023, Kocevski_2024, Harikane_2023, Kokorev_2024_census, Akins_2024, Taylor_2024}. Their redshift distribution peaks around $z \sim 5$, and they are broadly observed between $z \sim 4$ and $z \sim 8$, during $\sim 1$ Gyr of cosmic history. 
Studies are now detecting equivalents of the LRDs at lower redshifts (that is, $z < 4$), albeit at a significantly lower number density \citep{Ma_2025, Zhuang_2025, Rinaldi_2025, Billand_2025}, which decreases exponentially with decreasing redshift.

The LRDs are also compact, with an average effective radius of $\sim 150$ pc, and typically less than $\sim 300$ pc \citep{Baggen_2023}.
Inspired by these peculiarities, \cite{Pacucci_Loeb_2025} suggested that all three observed properties of the LRDs (i.e., abundance, compactness, and redshift distribution) are explained by their origin in low-spin halos.

Aside from the observed properties of the LRDs, which are peculiar enough, their inferred physical properties are puzzling. What powers their light? It is currently unclear whether the LRDs are primarily powered by an accreting supermassive black hole (SMBH) with typical masses $\sim 10^7 - 10^8 \Msun$ (see, e.g., \citealt{Maiolino_2023_new}) or by star formation in already massive galaxies, with stellar masses of $\sim 10^9 - 10^{11} \Msun$ (see, e.g., \citealt{Labbe_2023}). The V-shaped SEDs of the LRDs are characterized by an inversion that appears at the Balmer break \citep{Setton_2024}; moreover, several studies have shown that their SEDs can be explained with various AGN fractions (see, e.g., \citealt{Akins_2024, Durodola_2024}), which is the fraction of light generated by the central SMBH at a specific wavelength.

Most LRDs exhibit broad emission lines with widths of $\sim 1000-2000 \, \rm km \, s^{-1}$ \citep{Greene_2023}. 
This is typically associated with the presence of a central SMBH, which accelerates the swirling gas to high velocities. These signatures are widely used, both locally and in the distant Universe, to estimate the SMBH mass \citep{Greene_2005}.
However, several studies have also shown that the same signature can be produced by extremely massive and compact galaxies, in which the high core stellar density naturally produces extreme velocity dispersions \citep{Loeb_2024RNAAS, Baggen_2024}.

An additional complication arises from the lack of X-ray detections from the LRDs, even in deep stacking analyses \citep{Ananna_2024, Maiolino_2024_Xray}, although \cite{Yue_2024_Xray} claims a marginal detection. This is troublesome for the black hole hypothesis, as the presence of X-ray emission is generally heralded as a trademark for the presence of a SMBH.
However, several models explain the lack of X-ray emission while saving the black hole hypothesis.
For example, \cite{Maiolino_2024_Xray} suggest that very thick absorbing material, with a significant covering factor, would obscure the X-ray emission. \cite{Pacucci_Narayan_2024} propose that mildly super-Eddington accretion onto a slowly spinning SMBH would create an SED that is so steep and soft in the X-ray as to be effectively undetected by current X-ray observatories. The super-Eddington accretion hypothesis has also been studied by \cite{Lupi_2024}, \cite{Lambrides_2024}, and \cite{Madau_2024}.

The star-only interpretation is also challenged. In particular, the combination of two factors leads to a significant issue: (i) the requirement of stellar masses in the range of $10^9-10^{11} \Msun$ to reproduce the observed light, and (ii) their compactness, with effective radii $\sim 80-300$ pc. The resulting core stellar densities are immense, as shown by \cite{Baggen_2024} and \cite{Guia_2024}. In particular, the latter study shows that $35\%$ of a sample of 475 photometrically selected LRDs (from the COSMOS-Web and PRIMER surveys) exhibit core stellar densities that are higher than the highest stellar densities observed in various stellar systems in any redshift range, reaching densities as high as $\sim 10^9 \rm \Msun \, pc^{-3}$. These values are $>10$ times higher than the density required for runaway stellar collisions to occur \citep{2004Natur.428..724P,Ardi_2008, Fujii_2024}. A runaway process is one in which the growth of the central star occurs on a timescale shorter than the star's lifetime. Assuming the environmental properties estimated for the LRDs, the (average) stellar collision time \citep{BT_2008} is $\ll 1$ Myr, which also warrants that the runaway conditions are met. Runaway stellar collisions have long been proposed as a mechanism to form massive black holes (MBHs, \citealt{Sanders_1970, BR_1978, Rees_1984, Quinlan_1990}). Stellar collisions in the dense environments around central SMBHs, including the one in the Milky Way, have been extensively studied (e.g., \citealt{MB_2021, Balberg_2024}), and have been proposed as an effective mechanism for the formation of massive stars \citep{Rose_2023}.

In this study, we investigate this phenomenon.
Assuming the star-only interpretation, characterized by extreme core stellar densities, we ask whether the resulting stellar system remains stable or if runaway stellar collisions inevitably lead to the formation of a MBH in the core of the LRDs.
In Sec.~\ref{sec:methods}, we describe the three complementary approaches used to explore the dynamical environment of the LRDs, while Sec.~\ref{sec:results} presents our main findings. Finally, in Sec.~\ref{sec:disc_concl}, we discuss the broader implications of our results, proposing that the LRDs may represent the progenitors of the SMBHs observed in local galaxies.

\section{Methods}
\label{sec:methods}

To assess the dynamical consequences of the extreme stellar densities inferred in the cores of the LRDs, we adopt three complementary approaches.

First, we use a Fokker-Planck solver (Sec. \ref{subsec:FP}) to track the secular evolution of the stellar density profile, starting from initial conditions representative of a typical LRD, i.e., one with a core density comparable to the median of the inferred distribution \citep{Guia_2024}.
This method, along with the associated analysis of the relevant timescales, provides an enhanced understanding of the core dynamics.
Second, we develop a simplified analytical framework (Sec. \ref{subsec:analytical}) to model the growth of a central very massive star (VMS) through stellar collisions. This method assesses whether a central VMS will form via runaway collisions, and what mass it should reach before its growth is halted.
Third, and finally, we perform a direct $N$-body simulation of dense stellar clusters (Sec. \ref{subsec:Nbody}) to confirm the formation of a VMS and quantify its mass growth and the timescale over which this process occurs.

These complementary methods offer a consistent picture of rapid core collapse in ultracompact stellar systems, such as the LRDs. Their application to the LRD case is described in Sec. \ref{sec:results}.

\subsection{Fokker-Planck Analysis}
\label{subsec:FP}

The collisional time evolution of dense stellar systems can be described by the orbit-averaged Fokker-Planck equation, which models the cumulative effect of weak, uncorrelated two-body encounters as a diffusion process in the relevant phase space. This process can be seen as replacing the $N$ quantized bodies with a fluid of particles. 
By performing an ``orbit-average'', we effectively assume that changes in the distribution function (DF) caused by encounters between bodies are small over a single orbital period, because the relaxation time in systems with $N \gg 1$ is significantly longer than the crossing time.
\cite{BT_2008} provides an in-depth description of the Fokker-Planck method and its astrophysical applications.

The one-dimensional orbit-averaged Fokker–Planck equation, in the flux-conservative form \citep{BT_2008, Vasiliev_2017}, can be written as: 

\begin{equation}
\frac{\partial [f(E, t) \, g(E)]}{\partial t} = -\frac{\partial F(E, t)}{\partial E} \, ,
\end{equation}
where $f(E, t)$ is the DF in terms of energy and time, $g(E)$ is the density of states defined as $g(E) \equiv {\rm d} h(E)/{\rm d}E$ (where $h$ is the phase volume), and $F(E, t)$ is the flux in energy space. This equation describes the diffusion in energy space.

In our work, we employ the Fokker-Planck method via the publicly available code \texttt{PhaseFlow} \citep{Vasiliev_2017}, which implements a high-accuracy solver. This code relies on a reformulation of the general Fokker-Planck formalism, which writes the relevant equations in terms of the phase volume $h$. This term is defined as the volume of phase space enclosed within the hypersurface of a given energy $E$. 

Expressing the DF as $f(h, t)$ (i.e., in terms of phase volume instead of energy) has several advantages: the phase volume is an adiabatic invariant under slow changes in the potential, and a uniform grid in $\ln h$ efficiently resolves large dynamical ranges in energy.
In this formulation, the Fokker-Planck equation becomes:
\begin{equation}
\frac{\partial f(h, t)}{\partial t} = -\frac{\partial F(h, t)}{\partial h} + s(h, t) - \nu(h, t) \, f(h, t) \, ,
\end{equation}
where $s$ and $-\nu f$ are optional source and sink terms. The flux is given by:

\begin{equation}
F(h, t) = -D(h) \, \frac{\partial f}{\partial h} - A(h) \, f(h, t) \, ,
\label{eq:FP_eq}
\end{equation}
where $D(h)$ and $A(h)$ are the diffusion and advection coefficients, expressed in terms of phase volume. 

The DF is evolved on a logarithmically spaced phase-volume grid with $200$ points, ranging from $h = 10^{-20}$ to $h = 10^3$. We use the Chang \& Cooper method (i.e., method=0 in \texttt{PhaseFlow}), which ensures conservative flux integration across the grid. The timestep is adaptively controlled via an accuracy parameter $\epsilon = 10^{-4}$, which guarantees excellent energy conservation throughout the evolution.

The initial conditions for our run are based on typical conditions estimated at the core of LRDs. In particular, we choose the median stellar density calculated for a large sample of LRDs \citep{Baggen_2024, Guia_2024}: $\rho_\star \sim 10^4 \, \Msun \, \mathrm{pc}^{-3}$. Note that the latter study suggests that LRDs with core densities up to $\rho_\star \sim 10^9 \, \Msun \, \mathrm{pc}^{-3}$ may exist, assuming that the measured effective radii and stellar masses are correct. More details on the specific initial conditions are provided in Sec. \ref{subsec:Nbody_res}.

\subsection{Analytical Calculation}
\label{subsec:analytical}

We use an analytical model to estimate the growth of a central VMS via stellar collisions in a dense stellar system, such as the LRDs. This model was used and described in detail in \cite{Fujii_2024} to show that intermediate-mass black holes can form in local globular clusters. The overarching goal is to model the competition between mass gain from runaway stellar collisions and mass loss via stellar winds.

The VMS grows through repeated inelastic collisions between stars in the core of the stellar system. The average mass accretion rate from these collisions, $\dot{M}_{\rm col}$, is determined by: (i) the rate of collisions $\dot{N}_{\rm col}$, and (ii) the typical mass contributed per collision $\langle \delta m \rangle$. The accretion rate is then:

\begin{equation}
\dot{M}_{\rm col} = \dot{N}_{\rm col} \langle \delta m \rangle \, .
\end{equation}
In high-density systems, the collision rate is primarily governed by dynamically formed binaries, which rapidly harden and merge \citep{PZ_2002}. The rate of such collisions is estimated as:

\begin{equation}
\dot{N}_{\rm col} \approx 10^{-3} f_c \frac{N_{\rm cl}}{t_{\rm rh}} \, ,
\end{equation}
where $f_c \sim 0.8$ is the fraction of dynamically formed binaries that result in collisions, $N_{\rm cl}$ is the total number of stars, and $t_{\rm rh}$ is the half-mass relaxation time. The mean accreted mass per event is approximated as:

\begin{equation}
\langle \delta m \rangle \simeq \frac{4 \langle m \rangle \ln \Lambda}{t} \, t_{\rm rh} \, ,
\end{equation}
where $\langle m \rangle$ is the average stellar mass and $\ln \Lambda$ is the Coulomb logarithm. Combining the expressions, we obtain the accretion rate:

\begin{equation}
\dot{M}_{\rm col} \sim 4 \times 10^{-3} f_c \ln \Lambda \frac{N_{\rm cl} \langle m \rangle}{t} \, .
\end{equation}

Note that the Coulomb logarithm $\ln \Lambda$ is a number of the order of $\sim 10$ for large clusters.
Using $N_{\rm cl} \langle m \rangle = M_{\rm cl}$ (i.e., the total cluster mass) and expressing the star formation rate as $\dot{M}_{\rm SF} = M_{\rm cl} / t$ (where $t$ is the time elapsed since the formation of the cluster, of current mass $M_{\rm cl}$), we find $\dot{M}_{\rm col} \sim 0.04 f_c \dot{M}_{\rm SF}$, or: 
\begin{equation}
\dot{M}_{\rm acc} \approx 0.03 \, \dot{M}_{\rm SF} \, ,
\label{eq:accretion_rate}
\end{equation}
for $f_c = 0.8$, which is the empirical result found in \cite{Fujii_2024}. The mass accretion rate onto the central VMS is estimated as $\sim 3\%$ of the star formation rate in the stellar cluster. This result holds across a wide range of stellar densities and metallicities and is the outcome of mass segregation and binary-induced collisions in the evolution of a compact cluster.

To estimate the equilibrium mass of the VMS, interpreted as a balance between mass acquired by collisions and mass loss via stellar winds, we evolve its mass as a function of time via:

\begin{equation}
\frac{{\rm d}M_{\rm VMS}}{{\rm d}t} = \dot{M}_{\rm acc} - \dot{M}_{\rm wind}(M,Z) \, ,
\label{eq:mass_evolution}
\end{equation}
where $\dot{M}_{\rm wind}$ is the metallicity-dependent stellar winds mass loss rate \citep{Vink_2018}:

\begin{eqnarray}
\log_{10} \left( \frac{\dot{M}_{\rm wind}}{M_\odot \, \mathrm{yr}^{-1}} \right) &=& -9.13 + 2.1 \log_{10} \left( \frac{M_{\rm VMS}}{M_\odot} \right) \nonumber \\
&& + 0.74 \log_{10} \left( \frac{Z}{Z_\odot} \right) \, .
\label{eq:mass_loss}
\end{eqnarray}

Assuming that the metallicity $Z$ of the VMS remains constant, the mass loss via winds increases with the mass of the VMS, $M_{\rm VMS}$. Hence, its mass will increase until its derivative, ${\rm d}M_{\rm VMS}/{\rm d}t$, is zero.

\subsection{N-body Simulations}
\label{subsec:Nbody}

Finally, we conduct direct $N$-body simulations of a dense stellar system undergoing runaway stellar collisions. These simulations track individual stellar orbits and resolve close gravitational encounters with high precision, enabling a self-consistent treatment of the formation of a central VMS through successive mergers.

The initial model is constructed using \texttt{Mcluster} \citep{2011MNRAS.417.2300K}. In particular, we adopt a Plummer density profile \citep{1911MNRAS..71..460P} with $N = 10^6$ stars drawn from a Kroupa initial mass function (IMF, \citealt{2003ApJ...598.1076K}) with an upper- and lower-mass cut-off of $0.1 \Msun$ and $100 \Msun$. By adopting $0.2$ \,pc for the scale radius, the initial central stellar density is $10^7 \, \Msun \, \mathrm{pc}^{-3}$, representative of the inferred conditions in a high-density LRD \citep{Guia_2024} and consistent with the central density reached naturally after a relaxation time for the typical LRD, as shown by our Fokker-Planck analysis (see Sec. \ref{subsec:FP_res}). No central black hole, gas component, or external tidal field is included in our simulations. 

We use the code \texttt{PETAR} \citep{Wang_Iwasawa_2020, Fujii_2024}, a high-performance $N$-body integrator optimized for collisional stellar systems. \texttt{PETAR} employs a hybrid approach \citep{2011PASJ...63..881O} that couples a direct integration scheme for close encounters (based on the Hermite method) with a tree algorithm for long-range forces, achieving accuracy and scalability. The code includes detailed prescriptions for stellar evolution, mass loss, and binary formation, and can treat gravitational interactions involving $>10^6$ particles.

When two stars pass at a distance between their centers of mass that is smaller than their combined radii, they merge without mass loss (i.e., the so-called ``sticky-sphere'' approximation) and continue as a single object with updated mass and radius. 
This approximation is valid in the innermost region of our model, where the local velocity dispersion is $\lesssim 10 \, \rm km \, s^{-1}$ and stellar encounters are sub‐escape‐velocity collisions (see Sec. \ref{sec:results} and Fig. \ref{fig:FP_timescales}). Although the global velocity dispersion of the cluster can exceed $10^3$ km s$^{-1}$, the stellar collisions that lead to runaway growth occur in the very center, where the relative velocities are sufficiently low for the sticky‐sphere treatment to hold (see also \citealt{Fujii_2024}).

While stellar mergers can temporarily inflate the resulting object due to deposited orbital energy, detailed simulations (e.g., \citealt{Glebbeek_2009, Ramirez-Galeano_2025}) show that only a small fraction of the stellar mass is unbound during the process, as most of the energy is carried away by a limited amount of ejecta. Consequently, the merger products remain largely bound and return to hydrostatic equilibrium on short timescales. In practice, such transient inflation may even increase the effective collisional cross section, thereby facilitating rather than suppressing the runaway sequence. The stellar radii are calculated using SSE \citep{2002MNRAS.329..897H,2020A&A...639A..41B} included in \texttt{PeTar}. For VMS, we adopted the mass loss rate provided in Eq.~\ref{eq:mass_loss}. 

Additionally, in our simulation, we do not explicitly track the stellar spins. However, the adopted cluster model has no initial net rotation, and collisions occur with random orientations, so the net angular momentum of the merger products is expected to remain low.

The system is evolved for $\sim 1$ Myr, which is significantly longer than the central dynamical friction time for the stellar densities considered. A summary of the main properties of our $N$-body simulation is provided in Table \ref{tab:ICs}.

\begin{table}[ht]
\centering
\caption{Initial conditions of our $N$-body simulation.}
\begin{tabular}{lc}
\hline
\hline
\textbf{Parameter} & \textbf{Value} \\
\hline
\hline
Density profile & Plummer \\
Scale radius & 0.2 pc \\
Total mass & $5.78 \times 10^5 \, \Msun$ \\
Total particle number & $10^6$  \\
IMF (Kroupa) & $0.08$--$100 \, \Msun$ \\
\hline
\end{tabular}
\label{tab:ICs}
\end{table}

\section{Results}
\label{sec:results}

This Section applies the methodologies described above to investigate whether the physical conditions inferred for the LRDs trigger core collapse and the formation of a VMS via stellar collisions. 

\subsection{Fokker-Planck Analysis: Results}
\label{subsec:FP_res}

The Fokker-Planck simulation is initialized with a Plummer density profile \citep{1911MNRAS..71..460P} with total mass $M_\star = 10^{11} \Msun$ and scale radius $a = 100$ pc. This corresponds to an initial core density:  

\begin{equation}
\rho_0 = \frac{3}{4\pi}\,\frac{M_{\rm tot}}{a^3}
\approx 2.4\times10^4\, \Msun\,\mathrm{pc}^{-3}.
\end{equation}

Note that $\rho_\star \sim 10^{4-5} \, \Msun\,\mathrm{pc}^{-3}$ is the median core density estimated for LRDs systems in \cite{Guia_2024}. Hence, our current goal is to simulate the core dynamics of the typical LRD in the star-only interpretation.  

\begin{figure*}[ht!]
\includegraphics[width=0.49\textwidth]{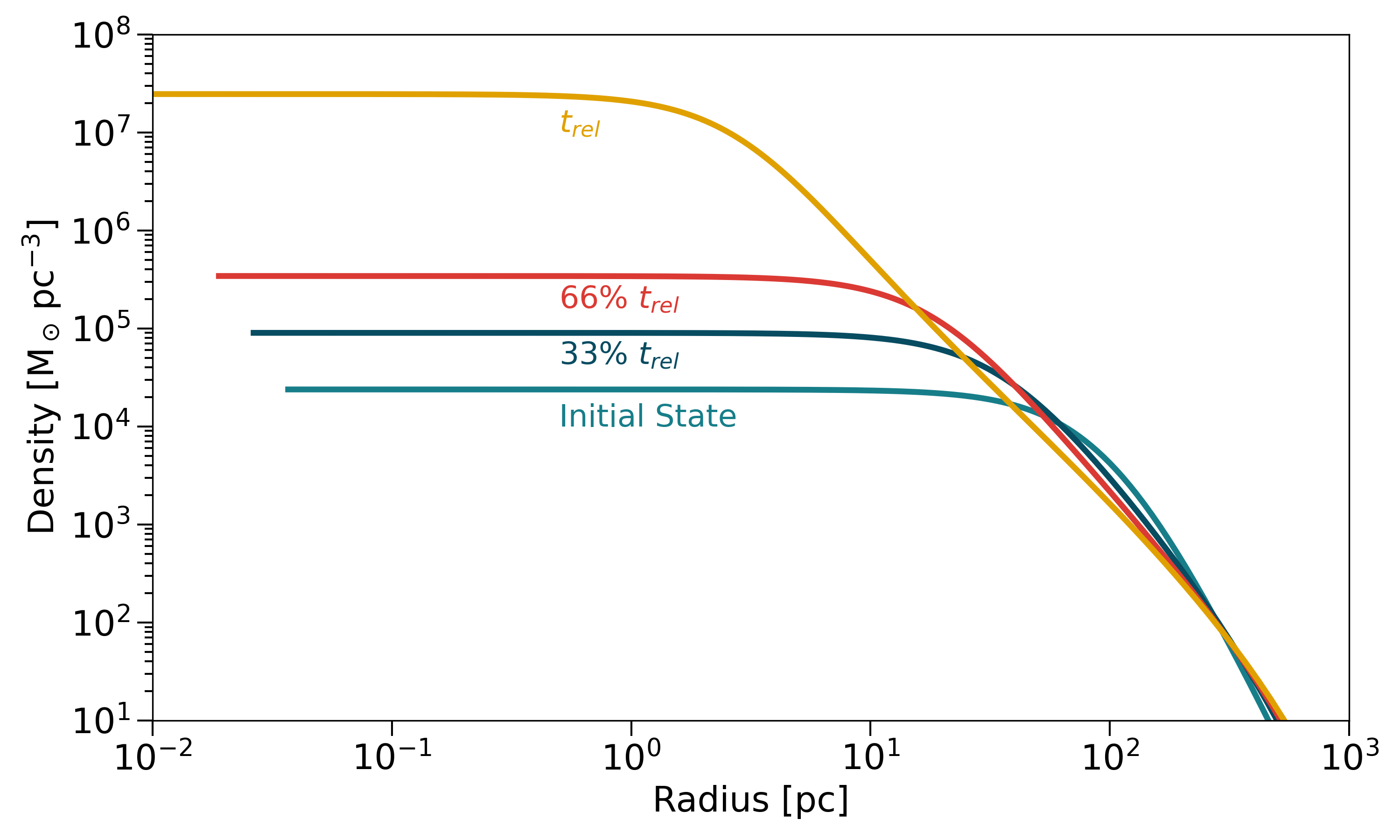}
\includegraphics[width=0.49\textwidth]{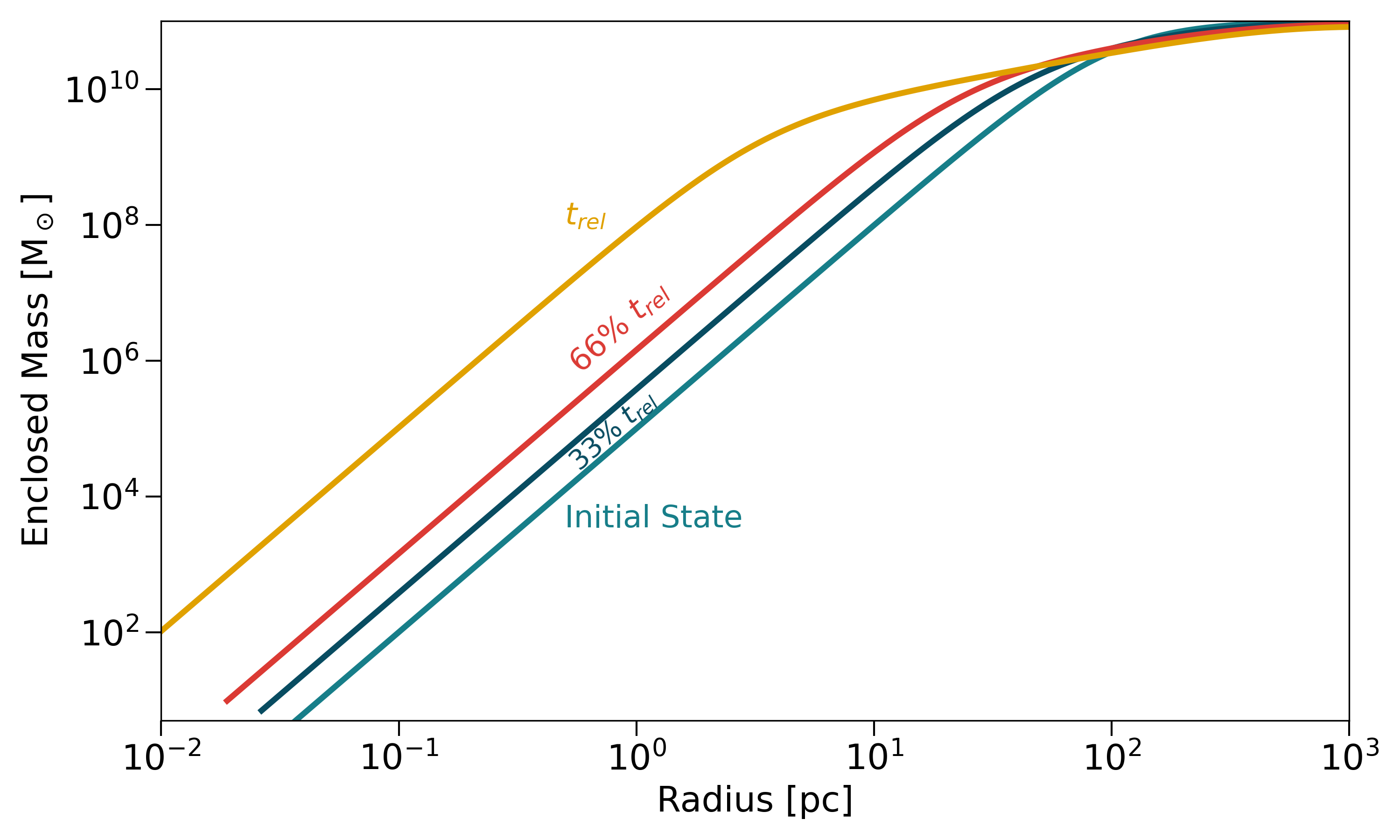}
\caption{\textit{Left:} Radial density profiles at initial, final, and two intermediate collapse times, labeled by fraction of the global relaxation time $t_{\rm rel}$.  \textit{Right:} Corresponding enclosed mass profiles.  The central parsec gains four orders of magnitude in mass, while the outer layer remains static.}
\label{fig:FP_density}
\end{figure*}

Figure~\ref{fig:FP_density} (left panel) displays the evolution of the density profile from the initial state through the estimated global relaxation time $t_{\rm rel}$, via two milestones at $0.33\, t_{\rm rel}$ and $0.66\, t_{\rm rel}$. The global relaxation time is $t_{\rm rel} \sim 10^7\;\mathrm{Myr}$ computed via:

\begin{equation}
t_{\rm rel}
= \biggl(\frac{0.1\,N}{\ln N}\biggr)\,\sqrt{\frac{r^3}{G\,M(r)}} \, ,
\end{equation}
where $N$ is the total number of bodies in the system \citep{BT_2008}. In the presence of a strongly collisional system, a better approximation of the evolution timescale for the cluster core is the dynamical friction time, or the relaxation time computed locally. For example, in our extremely dense cluster, the relaxation time in the innermost $\sim 0.5$ pc is only $\sim 1$ Myr.

As shown in Fig. \ref{fig:FP_density}, by $t_{\rm rel}$ the central density has risen by three orders of magnitude, reaching $>~10^7\, \Msun\,\mathrm{pc}^{-3}$; the external density profile remains almost unchanged.

The corresponding enclosed mass evolution during the global relaxation time (Fig.~\ref{fig:FP_density}, right panel) shows that, within $r=1$ pc, $M(r)$ grows from $\sim10^5 \Msun$ to $\sim10^8 \Msun$.  Outside $\sim50$ pc, the mass profile remains unchanged, confirming that the collapse is strictly an inner‐core phenomenon.

\begin{figure}[ht!]
\includegraphics[width=0.49\textwidth]{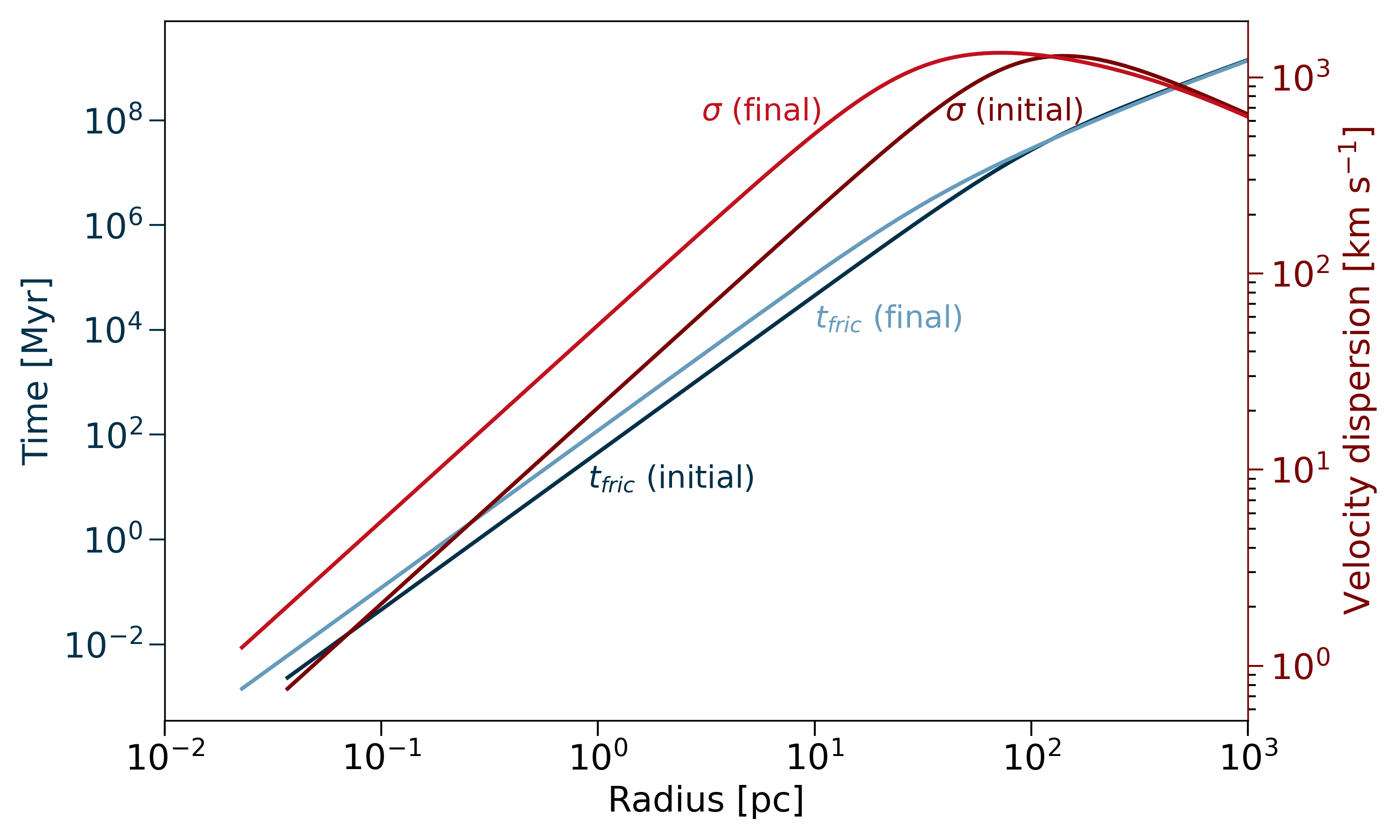}
\caption{Dynamical friction time and velocity dispersion before (dark) and after (light) core contraction. The dynamical friction time $t_{\rm fric}$ is calculated for a massive star of $10 \Msun$; in the central $0.1$ pc of the core, it falls to $< 0.1$ Myr, enabling rapid mass segregation. The red curves (right axis) show the one‐dimensional velocity dispersion $\sigma(r)$.}
\label{fig:FP_timescales}
\end{figure}

Figure~\ref{fig:FP_timescales} illustrates the key timescale for the local evolution of the system, as a function of radius, before (dark) and after (light) core contraction. The dynamical friction time \citep{BT_2008} for a massive star of $M_{10}= 10 \Msun$ (assuming all the other stars are Sun-like) is calculated as:
\begin{equation}
    t_{\mathrm{fric}}(r) = \frac{1.65 \, r^{2} \, \sigma(r)}{\ln\Lambda \; G \; M_{10}} \, .
\end{equation}
The dynamical friction time in the central $0.1$ pc drops to $<0.1$ Myr. Such a short time indicates that the most massive stars of the cluster segregate rapidly into the shrinking core, setting the stage for runaway collisions on timescales of $\sim 1$ Myr. Importantly, the typical lifetime for a massive star of $10 \Msun$ is $\approx 25-30$ Myr \citep{Schaller_1992}, which is significantly longer. Hence, we conclude that the runaway process occurs.

We also compute the one‐dimensional velocity dispersion, $\sigma(r) = \sqrt{G\, M(r)/r}$, which is shown as the red curve on the right axis of Fig.~\ref{fig:FP_timescales}. Overall, $\sigma(r)$ peaks at $>10^3\,$km\,s$^{-1}$ on spatial scales of $\sim 70-100$ pc, reflecting the deepened potential well. Velocity dispersions of $>10^3\,$km\,s$^{-1}$ were invoked by \cite{Loeb_2024RNAAS} and \cite{Baggen_2024} to explain the broad emission lines observed in many LRDs (see, e.g., \citealt{Greene_2023}), within the star-only interpretation.

In sum, the Fokker–Planck results quantify a slow contraction of the core and a rapid mass segregation of the heavy stellar component. While full core collapse requires a timescale longer than the age of the Universe, the fast dynamical friction of massive stars ensures that a high‐density nucleus can nonetheless form on cosmologically fast timescales, possibly leading to the formation of a VMS via stellar collisions.

\subsection{Analytical Calculation: Results}
\label{subsec:analytical_res}

\begin{figure*}[ht!]
\centering
\includegraphics[width=0.90\textwidth]{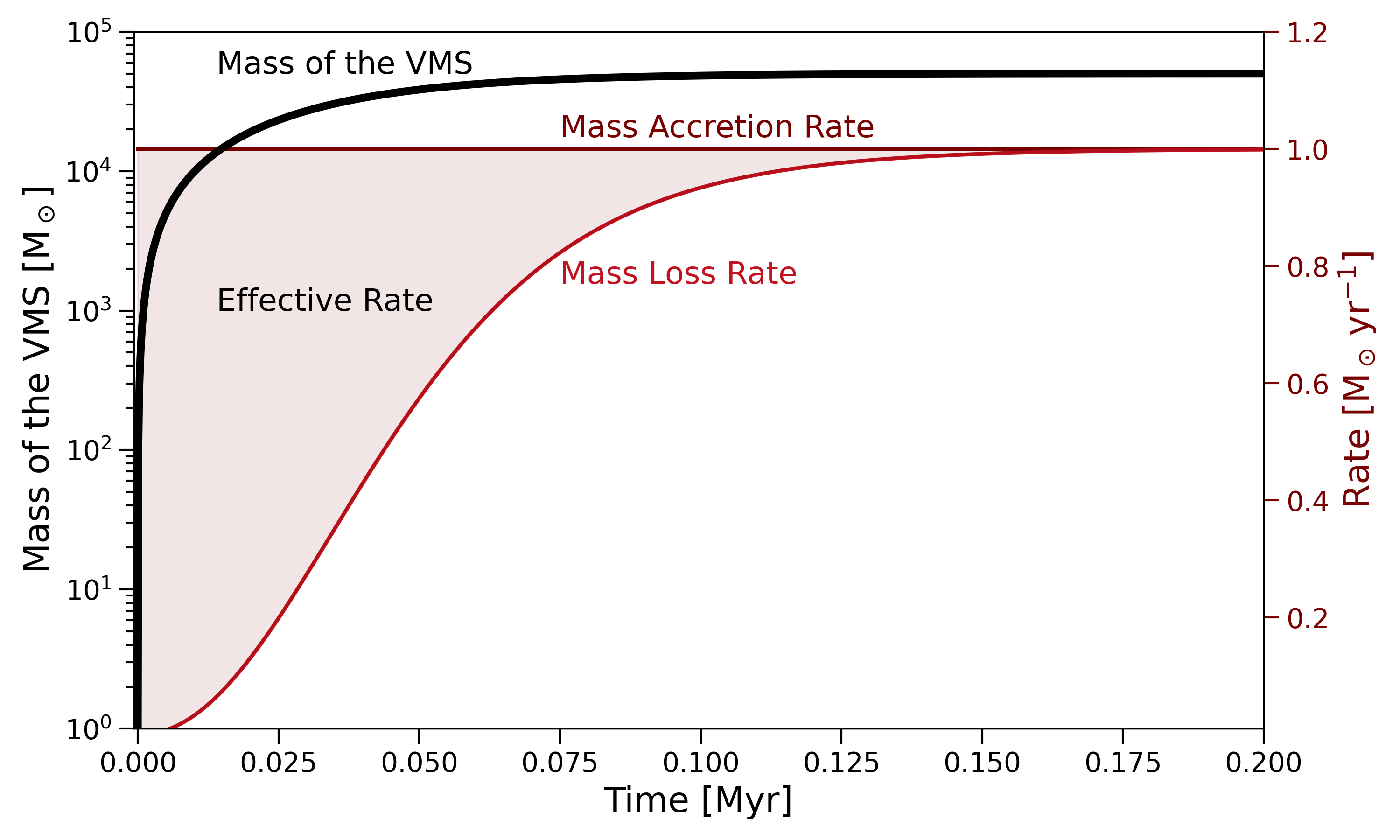}
\caption{Time evolution of the mass of the VMS (left axis) and the mass accretion and loss rates (right axis) onto it, calculated in our analytical model. The effective rate of growth of the VMS is given by the (shaded) difference between mass accretion and mass loss rates. The final mass, $M_{\rm VMS} = 5\times 10^4 \Msun$, is reached in a time comparable to the dynamical friction timescale in the central $0.1$ pc of the core.
\label{fig:analytical}}
\end{figure*}

We now evaluate whether a VMS can grow, via stellar collisions, rapidly enough before stellar winds limit its mass. Using the analytical framework described in Sec.~\ref{subsec:analytical}, we solve the mass evolution equation (Eq. \ref{eq:mass_evolution}) accounting for both a constant collisional accretion rate (Eq. \ref{eq:accretion_rate}) and a mass- and metallicity-dependent stellar wind loss term (Eq. \ref{eq:mass_loss}). 

To perform the calculation, we need to estimate the star formation rate and the metallicity of a typical LRD. Regarding the former, estimates for the LRDs vary widely. For example, \cite{Killi_2024} estimate values between $\sim 12 \, \Msun\, \mathrm{yr}^{-1}$ and $\sim 49 \, \Msun\, \mathrm{yr}^{-1}$. \cite{Wang_RUBIES_2024} obtain values between $\sim 11 \, \Msun\, \mathrm{yr}^{-1}$ and $\sim 30 \, \Msun\, \mathrm{yr}^{-1}$, although instantaneous values may reach $\sim 150 \, \Msun\, \mathrm{yr}^{-1}$. \cite{Labbe_2024} infer values in an ultraluminous LRD at $z=4.47$ that are significantly higher. Given such a range of estimates, we assume a conservative value of $\sim 30 \, \Msun\, \mathrm{yr}^{-1}$.
Regarding the metallicity, estimates from \cite{Killi_2024} and \cite{Wang_RUBIES_2024} provide values between $6\% \, \rm Z_\odot$ and $32\% \, \rm Z_\odot$, while \cite{Labbe_2024} provides a value of $10\% \, \rm Z_\odot$. Hence, we reasonably assume a reference value of $Z = 10\% \, \rm Z_\odot$. From the star formation rate, we calculate a constant accretion rate of $\dot{M}_{\rm acc} \sim 1 \, \Msun \, \mathrm{yr}^{-1}$.

The results for the time evolution of the VMS are displayed in Fig. \ref{fig:analytical}. The mass of the VMS grows rapidly, exceeding $10^4 \Msun$ within just $0.1$ Myr, which is consistent with the dynamical friction timescale in the central $0.1$ pc (see Sec. \ref{subsec:FP_res}). The effective mass accretion rate, shaded in Fig. \ref{fig:analytical}, is the difference between the mass accretion rate (due to stellar collisions) and the wind mass-loss rate. In our model, the mass accretion rate is steady at $\sim 1 \rm \Msun \, yr^{-1}$ (i.e., $3\%$ of the assumed star formation rate), while the mass accretion rate increases with the mass of the VMS (Eq. \ref{eq:mass_loss}). 

Accretion essentially stops once the effective rate reaches zero, as the VMS ejects via winds as much mass as it accretes from the infalling stars. This analysis confirms that stellar winds are not sufficient to counteract collisional growth for a time comparable to the dynamical friction timescale of the core. In this analysis, the final mass of the VMS is $M_{\rm VMS} = 5\times 10^4 \Msun$.
These findings are consistent with both theoretical expectations and numerical simulations of runaway stellar collisions in high-density systems \citep[e.g.,][]{Fujii_2024, Vergara_2025}.

\subsection{$N$-body Simulations: Results}
\label{subsec:Nbody_res}

\begin{figure*}[ht!]
\centering
\includegraphics[width=0.90\textwidth]{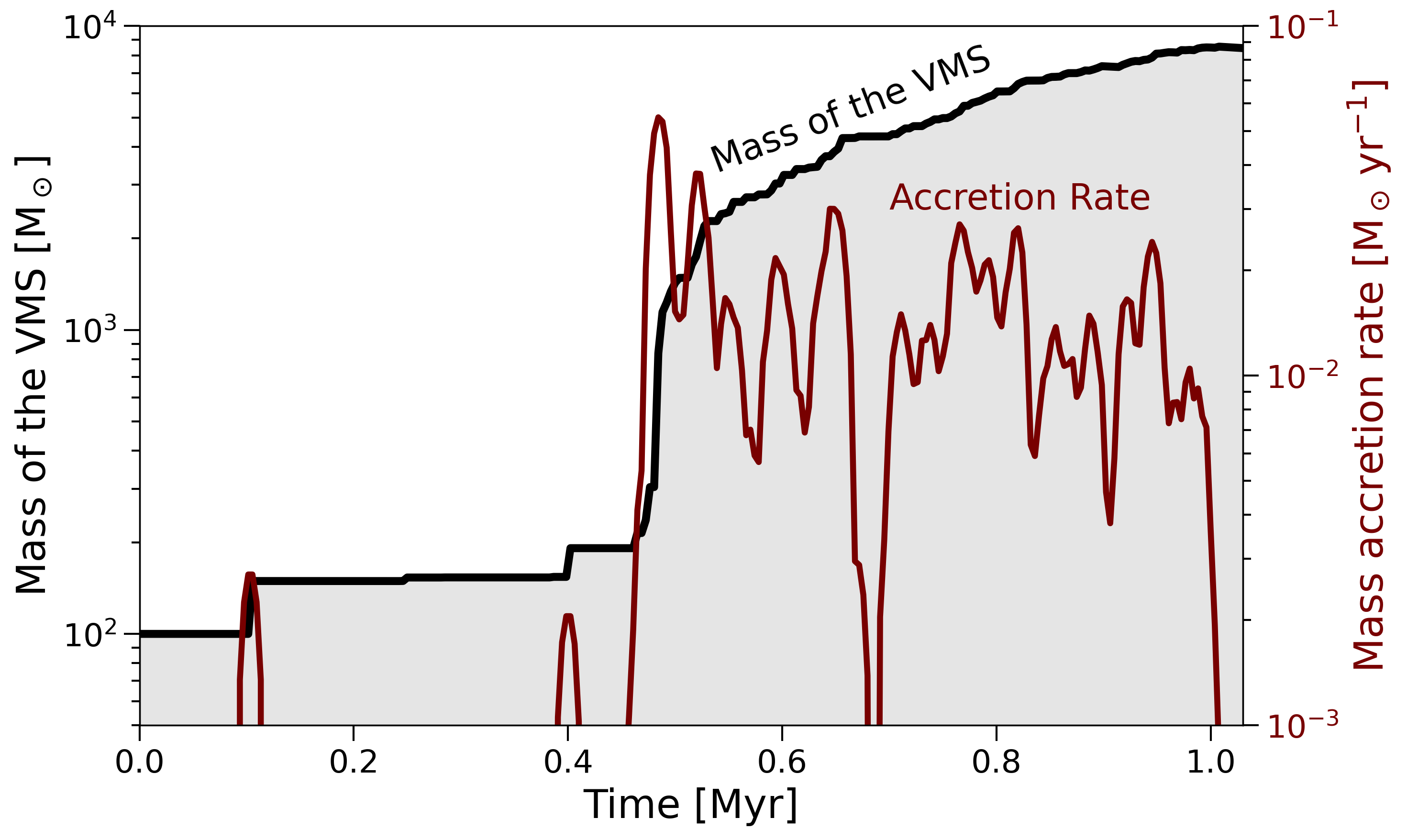}
\caption{Time evolution of the VMS mass in the direct $N$-body simulation. The system contains $10^6$ stars drawn from a Kroupa IMF and is initialized with a central stellar density of $10^7 \, \Msun \, \mathrm{pc}^{-3}$. The early phase shows discrete merger steps, followed by a runaway growth episode after $\sim 0.5$~Myr that drives the VMS to $9000 \Msun$ within $\sim 1$ Myr. The shaded region indicates the cumulative mass growth. The instantaneous mass accretion rate is also displayed in red, on the right vertical axis.}
\label{fig:Nbody}
\end{figure*}

Figure~\ref{fig:Nbody} shows the time evolution of VMS mass in our direct $N$-body simulation, whose technical details are described in Sec. \ref{subsec:Nbody}. The instantaneous mass accretion rate is also displayed.

For the first $\sim 0.45$ Myr, the VMS mass remains nearly constant, aside from a few discrete merger events, indicating a quasi-stable configuration as massive stars gradually sink toward the center via dynamical friction. 

A dramatic change occurs at $\sim 0.5$~Myr, when mass segregation has concentrated enough massive stars in the core to push the system past the critical density for runaway collisions (see a similar analysis in \citealt{Fujii_2024}). In the next $\sim 0.1$~Myr, the VMS mass grows steeply by more than an order of magnitude through successive mergers, eventually reaching $\sim 10^4\, \Msun$. After this runaway phase, the growth rate slows, and the VMS mass asymptotes toward its final value of $9000\, \Msun$ by the end of the simulation. 

This particle-based simulation, with no imposed accretion prescription or feedback, clearly displays an initial period of mild evolution followed by a rapid, centrally driven growth of a dominant object. The $N$-body results confirm that, under LRD-like initial conditions, runaway collisions in the stellar core are a robust pathway to forming a massive central object on timescales comparable to the core's dynamical friction time. Such a VMS is a plausible MBH progenitor, as we shall describe in Sec. \ref{subsec:MBH_formation}.

\subsection{Bracketing the Final VMS Mass}
\label{subsec:comparison}

The analytical calculation led to a final mass $M_{\rm VMS} = 5\times 10^4\Msun$ (Sec. \ref{subsec:analytical_res}), while the N-body simulation to $M_{\rm VMS} = 9\times 10^3\Msun$ (Sec. \ref{subsec:Nbody_res}). We now detail why these values are consistent with one another, and allow us to bracket the typical range of VMS masses that we expect to form in a LRD.

The mass growth of the VMS is ultimately a balance between two competing forces: (i) the collisional mass accretion rate, and (ii) the stellar wind mass loss rate. As described in Eq. \ref{eq:mass_evolution}, $M_{\rm VMS}$ stops evolving once the stellar wind mass loss rate reaches the value of the collisional accretion rate:
\begin{equation}
    \frac{{\rm d} M_{\rm VMS}}{{\rm d}t} = 0 \iff \dot{M}_{\rm acc} = \dot{M}_{\rm wind} \,.
\end{equation}

While the collisional accretion rate is related to the stellar environment (e.g., its density), the wind mass loss rate depends on $M_{\rm VMS}$, as in $\dot{M}_{\rm wind} \propto M_{\rm VMS}^{2.1}$ (see Eq. \ref{eq:mass_loss}): more massive stars have stronger winds.

In the analytical calculation, the collisional accretion rate was set to $1 \, \Msun \, \rm yr^{-1}$, while in the N-body simulation, its time-averaged value is $0.03 \, \Msun \, \rm yr^{-1}$ (see Fig. \ref{fig:Nbody}). Hence, the analytical calculation is characterized by an accretion rate that is $\sim 33$ times higher. The VMS reaches an equilibrium mass when $\dot{M}_{\rm acc} = \dot{M}_{\rm wind}$, which requires that $\dot{M}_{\rm wind}$ for the analytical calculation be also $\sim 33$ times higher. This condition is achieved with a final $M_{\rm VMS}$ that is $\sqrt[2.1]{33} \approx 5.3$ times higher, which is what we find.

To conclude, the final VMS masses reached in the analytical and $N$-body calculations are consistent with each other, and are the equilibrium masses indicative of different environmental conditions. The $N$-body simulation is a purely gravitational experiment, while the analytical calculation assesses the effect of the presence of gas (via the star formation rate), which increases the collisional accretion rate.
Hence, we argue that the analytical calculation and the $N$-body simulation bracket the reasonable mass range that we expect for the VMS: $9\times 10^3 < M_{\rm VMS} \, [\Msun] < 5\times 10^4$.

\subsection{Massive Black Hole Formation}
\label{subsec:MBH_formation}

Once the phase of runaway stellar collisions ends, the subsequent evolution of the VMS naturally leads to the formation of a MBH. 

In the early stages, the VMS grows efficiently by accreting massive stars delivered via mass segregation. As shown in Fig. \ref{fig:Nbody}, the average mass accretion rate onto the VMS is $\sim 10^{-2} \Msun \rm \, yr^{-1}$, which supports the fact that the VMS increases its mass by $\sim 10^4 \Msun$ in $1$ Myr.

However, the accretion rate onto the VMS drops sharply after $\sim 1$ Myr, which is the dynamical friction timescale of the central $0.1-1$ pc region of the stellar cluster (see Sec. \ref{subsec:FP_res}). This is because mass segregation has depleted the core of easily-accretable high-mass stars (i.e., the heaviest stars in the Kroupa IMF distribution).

Without continued mass input, the VMS contracts quasi-statically on its Kelvin-Helmholtz timescale, radiating away its gravitational binding energy at approximately the Eddington luminosity \citep{Hosokawa_2013, Begelman_2008}. Unlike the Hayashi track followed by low-mass, fully convective protostars, the quasi-stellar structure we investigate here resembles a radiation-pressure-dominated supergiant protostar, with a compact contracting core and an inflated envelope \citep{Begelman_2008}. 

The Kelvin-Helmholtz timescale is:
\begin{equation}
t_{\rm KH} \;\equiv\; \frac{G M_{\star}^2}{R_\star L_\star} \, ,
\end{equation}
where $M_\star$ is the stellar mass of the VMS, $R_\star$ its radius, and $L_\star\simeq L_{\rm Edd}$ the luminosity. For our VMS with $M_\star \sim 10^4 \Msun$, $R_\star \sim 10^3 \, \rm R_\odot$, and $L_\star \sim 4\times 10^8 \, \rm L_\odot$, we find $t_{\rm KH} \;\approx\; 8 \times 10^3\ {\rm yr}$.

Because the measured accretion rates in the $N$-body simulation is $\sim 10^{-2}\, \Msun \,{\rm yr^{-1}}$ (Fig.~\ref{fig:Nbody}), well below the $\gtrsim 0.1\, \Msun \,{\rm yr^{-1}}$ required to sustain a permanently bloated supergiant protostar \citep{Begelman_2008}, the VMS is expected to contract efficiently. In the analytical calculation, we predict a higher collisional accretion rate (i.e., $\sim 1 \rm \Msun \, yr^{-1}$), which would sustain the supergiant protostar phase. However, this large collisional rate cannot be maintained for long. Collisions are primarily driven by massive stars, which are characterized by the highest cross sections; once they have migrated to the core, the collisional cross-section craters, and accretion effectively stops.

As contraction proceeds, the stellar radius approaches the general relativistic instability threshold. \citet{Baumgarte_Shapiro_1999} showed that a rotating VMS at the onset of collapse has an equatorial radius of order
\begin{equation}
R_{\rm GR} \;\simeq\; 640 \,\frac{G M_\star}{c^2},
\end{equation}
together with a dimensionless spin parameter $a = cJ/GM_\star^2 \approx 0.97$, where $J$ is the angular momentum; i.e., the star is near the mass-shedding limit. Once $R_\star \lesssim R_{\rm GR}$, the star becomes unstable to quasi-radial collapse \citep{Chandrasekhar_1964, Baumgarte_Shapiro_1999, Shibata_Shapiro_2002, Begelman_2006, Begelman_2008, Montero_2012}. 

The subsequent collapse occurs on the dynamical timescale:
\begin{equation}
t_{\rm dyn} = \sqrt{\frac{R_{\rm GR}^{3}}{G M_\star}}
          \;\approx\; 1.6\times 10^{4}\,\frac{GM_\star}{c^3}
          \;\approx\; 8\times10^{2}\ {\rm s} \, ,
\end{equation}
i.e., about $13$ minutes. By comparison, nuclear and thermal timescales at this stage remain orders of magnitude longer:
\begin{equation}
t_{\rm dyn} \ll t_{\rm KH} \sim 8 \times 10^{3}\,{\rm yr} \ll t_{\rm nuc} \gtrsim 10^{5}\,{\rm yr} \, .
\end{equation}
Hence, we conclude that nuclear burning and radiative processes cannot intervene on the collapse timescale: the stellar structure cannot adapt fast enough to the quasi-instantaneous general-relativistic collapse.

General relativistic hydrodynamic simulations confirm that $\sim 90\%$ of the VMS mass collapses directly into the black hole \citep{Shibata_Shapiro_2002}, while $\sim 10\%$ is lost due to neutrino emission \citep{Fryer_2012}. The precise remnant mass depends weakly on internal structure and rotation; however, for $M_\star \gtrsim 10^{4}\, \Msun$, stellar winds are unable to erode a significant fraction of the envelope before collapse, particularly at sub-solar metallicities \citep{Heger_2003, Glebbeek_2009, Vink_2018}. Consequently, the final black hole mass is expected to be close to that of the progenitor VMS.

Hence, we conclude that the extreme stellar densities implied within the cores of the LRDs (in the star-only interpretation) inevitably lead to the formation of a MBH, with a mass $\Mblack \sim 10^4 \Msun$.
If their observed compactness and stellar masses reflect true stellar densities, the LRDs are ideal environments for producing MBHs via dynamical channels. 

This runaway merger mechanism has long been proposed as one of the most viable mechanisms for forming intermediate-mass or even heavy black hole seeds in the very early Universe i.e., at $z \gtrsim 20$ \citep{BR_1978, Rees_1984, Inayoshi_review_2019}. In the case of the LRDs, this mechanism could extend to significantly more recent cosmic times, down to $z \sim 4$.

\section{Discussion and Conclusions}
\label{sec:disc_concl}

The discovery of the LRDs has potentially revealed a ``new kind'' of galactic population in the early Universe, with small effective radii and large stellar masses. Their power source remains uncertain: some may be accreting massive black holes \citep{Maiolino_2023_new, Kocevski_2024}, while others could be dominated by intense star formation \citep{Labbe_2023}. In the star-only scenario, the combination of large stellar masses and small radii implies typical central densities of $\rho_\star \sim 10^{4-5}\,\Msun\,\mathrm{pc}^{-3}$, and up to $\rho_\star \sim 10^9\,\Msun\,\mathrm{pc}^{-3}$ \citep{Guia_2024}, well into the regime where runaway stellar collisions occur.

Motivated by these observations, we studied the dynamical consequences of these extreme stellar densities, with three complementary approaches: a Fokker-Planck analysis, an analytical model, and a direct $N$-body simulation. Our main findings are summarized as follows:

\begin{itemize}
\item In typical LRDs ($\rho_\star \sim 10^{4-5}\,\Msun\,\mathrm{pc}^{-3}$), the dynamical friction time for $10\,\Msun$ stars in the central $0.1$ pc is short ($\lesssim 0.1$ Myr), allowing rapid mass segregation and the buildup of a dense nucleus.
\item At higher stellar densities ($\rho_\star \sim 10^7\,\Msun\,\mathrm{pc}^{-3}$), runaway stellar collisions begin to occur within $\sim~0.5$ Myr, assembling a VMS with a final mass in the range $9\times 10^3 \Msun < M_{\rm VMS} [\Msun] < 5\times 10^4$ in $\lesssim 1$ Myr.
\item Once the supply of massive stars is depleted, the VMS contracts on a Kelvin-Helmholtz timescale ($\sim 8000$ yr) and undergoes general relativistic collapse, producing a massive black hole with a mass comparable to that of the progenitor VMS.
\item This process leads to the formation of heavy black hole seeds of mass $\Mblack \sim 10^4 \Msun$ down to $z \sim 4$.
\end{itemize}

The above conclusions depend on the stellar densities inferred for the LRDs. Stellar mass estimates are subject to uncertainties in spectral modeling and AGN contamination (see, e.g., \citealt{Rinaldi_2025}). While some studies (e.g., \citealt{Berger_2025}, although for brighter AGN) suggest that the stellar masses are generally robust to within $\lesssim 0.3$ dex, larger errors cannot be excluded. If the true stellar masses are much lower, the corresponding densities, and the likelihood of runaway collisions, would be reduced.

Our study focuses on the dynamical consequences of high stellar densities; however, these densities may also explain other observed features. For example, \cite{Loeb_2024RNAAS} and \cite{Baggen_2024} have proposed that the large velocity dispersions ($\sigma \gtrsim 10^3$ km s$^{-1}$) implied by such compact systems would naturally produce the broad lines seen in many LRDs, without invoking AGN activity. These large velocity dispersions are confirmed also in our analysis (see Fig. \ref{fig:FP_timescales}), at spatial scales of $\sim 100$ pc. In this view, the broad lines trace the gravitational potential of the dense stellar core, consistent with the measured widths.

In addition, the formation of $\sim 10^4\,\Msun$ MBHs at the center of the LRDs may also support models in which the observed light is powered by tidal disruption events (TDEs). In particular, \cite{Bellovary_2025} proposed that the LRDs could be explained by dense stellar systems hosting an intermediate-mass black hole that produces recurring TDEs as stars are accreted. The collapse of the central VMS into a MBH leaves most of the surrounding, extremely dense stellar distribution intact: only a small fraction of the initial stellar mass is locked into the MBH. Hence, a substantial stellar reservoir remains available, leading to frequent stellar disruptions and, thus, sustained accretion. This process could naturally power the observed emission in some LRDs while contributing to the further, rapid growth of the MBH.

The runaway stellar collision pathway explored in this study is distinct from the widely studied direct collapse black hole (DCBH, \citealt{Loeb_Rasio_1994, Bromm_Loeb_2003, Lodato_Natarajan_2006, Begelman_2006}) hypothesis, which operates in pristine halos at $z \gtrsim 20-30$ \citep{BL01}. 

A MBH formed in the prototypical LRD at $z \sim 5$ (i.e., at the peak of their redshift distribution) cannot explain the $\Mblack \gtrsim 10^9\,\Msun$ quasars discovered at $z \sim 6-7$ \citep{Fan_2001, Fan_2003, Mortlock_2011, Wu_2015, Banados_2018, Wang_2021_quasar}, which are used to constrain seed models \citep{Pacucci_2022_search}.
The prototypical $z \sim 5$ LRD can produce heavy seeds at later epochs, potentially seeding the SMBHs of more recent galaxies. 

However, as pointed out by \cite{Pacucci_Loeb_2025} and displayed visually by \cite{Rinaldi_2025}, it is possible that the lack of numerous detections of LRDs at $z \gtrsim 7$ is an observational bias, due to their low surface brightness. Indeed, \cite{Taylor_2025_z9} have recently detected a LRD with broad emission lines at $z \approx 9.288$. If the population of LRDs extends to higher redshifts, they are clear candidates to seed also the population of $\Mblack \gtrsim 10^9\,\Msun$ quasars at $z \sim 6-7$.

A key advantage over DCBHs is that LRDs are far more common than optimistic predictions for DCBH number densities.
Typical number densities of LRDs are $10^{-4}-10^{-5} \, \rm cMpc^{-3}$ \citep{Kocevski_2024}, while DCBH formation models produce seeds at number densities of $10^{-6}-10^{-8} \rm cMpc^{-3}$ \citep{Dijkstra_2008, Dijkstra_2014}. In fact, the formation of heavy seeds in the LRDs does not require special conditions such as metal-free gas or a strong Lyman-Werner background \citep{Inayoshi_review_2019}.

Hence, if the inferred stellar densities are confirmed, the LRDs could represent a widespread and efficient nursery for the formation of heavy seeds in the later stages of cosmic history.

\begin{acknowledgments}
We thank the referee for providing insightful comments on the paper. F.P. acknowledges support from a Clay Fellowship administered by the Smithsonian Astrophysical Observatory. This work was also supported by the Black Hole Initiative at Harvard University, which is funded by grants from the John Templeton Foundation and the Gordon and Betty Moore Foundation. 
\end{acknowledgments}




\bibliography{ms}{}
\bibliographystyle{aasjournal}



\end{document}